\documentclass[10pt,reqno]{article}
\usepackage{amsmath,amssymb,amsthm}
\usepackage{esint}
\usepackage{bm}
\usepackage{url}
\usepackage{subfigure}
\usepackage{graphicx,color}
\usepackage{algorithm}
\usepackage{algpseudocode}
\usepackage{algorithmicx}
\usepackage{hyperref}
\usepackage{float}
\usepackage{booktabs,multirow}
\usepackage{hhline}
\usepackage{setspace}
\usepackage{subfigure}
\usepackage{xcolor}
\usepackage{lipsum}
\usepackage{wasysym}  
\usepackage{caption}
\usepackage{hyphsubst}
\usepackage{filecontents}
\usepackage{hyperref}

\hypersetup{colorlinks,citecolor=blue,linkcolor=blue, urlcolor=blue}
\usepackage{pdflscape}
\usepackage{breakurl}
\newcommand{\RR}{\mathbb{R}}

\def\bpsi{{\boldsymbol{\psi}}}
\def\bgamma{{\boldsymbol{\gamma}}}
\def\btheta{{\boldsymbol{\theta}}}
\newcommand{\ds}{\displaystyle}

\DeclareMathAlphabet{\itbf}{OML}{cmm}{b}{it}

\newcommand{\pathfigures}{Figures/}
\graphicspath{{\pathfigures}}

\date{\today} 
\begin{document}
	
	\title{Anticancer Peptides Classification using Kernel Sparse Representation Classifier
	}
	
	\author{
		Ehtisham Fazal\footnotemark[1]
		\and
		Muhammad Sohail Ibrahim\footnotemark[2] 
		\and
		Seongyong Park\footnotemark[3]
		\and
		Imran Naseem\footnotemark[4]~\footnotemark[5]~\footnotemark[6]
		\and 
		Abdul Wahab\footnotemark[7] 
	}
	\maketitle
	
	\renewcommand{\thefootnote}{\fnsymbol{footnote}}
	\footnotetext[1]{Lambda Theta, Karachi, Pakistan (email: ehtisham@lambdatheta.com).}
	\footnotetext[2]{College of Electrical Engineering, Zhejiang University, Hangzhou, China (e-mail: msohail@zju.edu.cn)}
	\footnotetext[3]{Asan Medical Center, University of Ulsan, College of Medicine, Department of Anesthesiology and Pain Medicine, 88 Olympic-ro 43-gil, Songpa-Gu, Seoul, 05505, South Korea (e-mail: sypark0215@amc.seoul.kr)}
	\footnotetext[4]{School of Electrical, Electronic and Computer Engineering, The University of Western Australia, Crawley 6009, Australia (e-mail: imran.naseem@uwa.edu.au)}
	\footnotetext[5]{College of Engineering, Karachi Institute of Economics and Technology, Korangi Creek, Karachi, Pakistan}
	\footnotetext[6]{Research and Development, Love for Data, Karachi, Pakistan}
	\footnotetext[7]{Department of Mathematics, School of Sciences and Humanities, Nazarbayev University, 53, Kabanbay Batyr Avenue, 010000, Astana, Kazakhstan (abdul.wahab@nu.edu.kz)}
	\renewcommand{\thefootnote}{\arabic{footnote}}
\begin{abstract} 
Cancer is one of the most challenging diseases because of its complexity, variability, and diversity of causes. It has been one of the major research topics over the past decades, yet it is still poorly understood. To this end, multifaceted therapeutic frameworks are indispensable. \emph{Anticancer peptides} (ACPs) are the most promising treatment option, but their large-scale identification and synthesis require reliable prediction methods, which is still a problem. In this paper, we present an intuitive classification strategy that differs from the traditional \emph{black box} method and is based on the well-known statistical theory of \emph{sparse-representation classification} (SRC). 
Specifically, we create over-complete dictionary matrices by embedding the \emph{composition of the K-spaced amino acid pairs} (CKSAAP).
Unlike the traditional SRC frameworks, we use an efficient \emph{matching pursuit} solver instead of the computationally expensive \emph{basis pursuit} solver in this strategy. Furthermore, the \emph{kernel principal component analysis} (KPCA) is employed to cope with non-linearity and dimension reduction of the feature space whereas the \emph{synthetic minority oversampling technique} (SMOTE) is used to balance the dictionary. The proposed method is evaluated on two benchmark datasets for well-known statistical parameters and is found to outperform the existing methods. The results show the highest sensitivity with the most balanced accuracy, which might be beneficial in understanding structural and chemical aspects and developing new ACPs. The Google-Colab implementation of the proposed method is available at the author's GitHub page (\href{https://github.com/ehtisham-Fazal/ACP-Kernel-SRC}{https://github.com/ehtisham-fazal/ACP-Kernel-SRC}).
\end{abstract}


\section{Introduction}\label{sec:introduction}
{A}{ccording} to the \emph{global cancer statistics 2020}, \cite{sung2021global}, cancer is one of the leading causes of mortality worldwide.
It is a diversified group of numerous complicated diseases, rather than a single one, marked by uncontrolled cell growth and a propensity to rapidly spread or infiltrate other body parts. Cancer's inherent complexity and heterogeneity have proven to be significant barriers to the development of effective anticancer therapies \cite{basith2017expediting}. Cancer can be treated with conventional clinical methods such as surgery, radiation, and chemotherapy, but these methods have drawbacks that can be painful for patients \cite{kaur2022data}. Although the aforementioned conventional methods deliver positive outcomes, they can also have some substantial adverse effects, including myelosuppression, cardiac toxicity, and gastrointestinal damage \cite{basak2021comparison}.

The discovery of anticancer peptides (ACPs) has transformed the paradigm for treating cancer. The ACPs can interact with the anionic cellular components of cancer cells and repair them selectively without harming normal or healthy cells in the body. This amazing feature of the ACPs is vital for therapeutic strategies. The ACPs are typically composed of $5$ to $50$ amino acids that are often synthesized using antimicrobial peptides (AMPs), many of which have cationic characteristics. These features have resulted in the development of novel alternative cancer therapies. 

The biggest challenge with the ACPs is distinguishing them from other synthetic or natural peptides \cite{yi2019acp}. Researchers employ a variety of approaches to identify the ACPs \cite{tyagi2013silico}. Although the experimental procedures are gold-standard methods, they are costly and time-consuming, and hence, unsuitable for large-scale searches for prospective ACP candidates. As a result, alternative methodologies for identifying APCs are desired. 

Technical advances in artificial intelligence (AI) have substantiated that it is a powerful tool for dealing with incredibly complex situations \cite{atif2022multi}. Many studies have used machine learning models to predict proteins and classify peptide sequences; see, for instance,  \cite{khan2018rafp,usman2020afp,park2020e3,al2021ecm,usman2021aop,chen2018ifeature}. Even for the ACPs alone, there are several \emph{in silico} approaches for identifying new ACPs. For instance,  Tyagi et al. have proposed a \emph{support vector machine} (SVM)-based classification algorithm in \cite{tyagi2013silico}. Another study, \cite{hajisharifi2014predicting}, employed Chou's \emph{pseudo amino acid composition} to predict the ACPs and tested their mutagenicity using the \emph{Ames test}. Generalized chaos game representation methods \cite{ge2019identifying}, deep learning-based short-term memory models \cite{yi2019acp}, ensemble learning models \cite{ge2020enacp}, augmentation strategies for improved classification performance \cite{chen2021acp}, and \emph{ETree classifiers}-based on \emph{amino acid composition} (AAC) \cite{agrawal2021anticp} are examples of alternative approaches.

Although existing machine learning techniques have some advantages for ACP prediction, there is still a need for improvement. For instance, deep learning models provide cutting-edge performance, but their \emph{black-box} nature obscures the classification judgment. A relatively simple model, on the other hand, may not provide appropriate classification accuracy. To that aim, the \emph{sparse-representation classification} (SRC) method provides a great balance, where constrained optimization is a proven method for explainable sparse modeling \cite{naseem2017ecmsrc,li2023multi,wright2008robust,hofmann2008kernel}. In the SRC, a test sample may be reconstructed using a linear combination of dictionary items with sparse weights under the basic principle \cite{usman2022afp,naseem2010sparse,bengio2013representation}.
The SRC is a non-parametric learning approach in which the magnitude of a sparse vector corresponds to the contribution of the dictionary atoms   \cite{elad2010sparse,li2013simultaneous}. 

Various sparse vector combinations can be used to tackle optimization and ill-posed problems. \emph{basis-pursuit} (BP) \cite{chen1994basis}, \emph{orthogonal-matching-pursuit} (OMP) \cite{pati1993orthogonal}, and \emph{matching-pursuit} (MP) \cite{gharavi1998fast} are some prominent methods for the SRC. These strategies employ $l_1$-norm regularization to relax $l_0$-norm rigid sparsity constraint, allowing gradient estimation from continuous error surfaces \cite{zhang2010sparseness,mandal2016employing}.  The BP furnishes the most sparse solution, but its computing cost grows exponentially. The MP is faster than the BP and OMP, although its sparsity is not guaranteed. 

Aside from the optimization approach, the efficacy of the \emph{over-complete dictionary} (OCD) is the most important feature for the construction of an SRC model. In this regard, Zhang et al. proposed a kernel SRC \cite{zhang2011kernel}. The kernel mapping converts the nonlinear relationship between different atoms (samples in OCD) to a linear relationship, allowing the classification of even more complex patterns \cite{atif2022multi,khan2017novel,zhang2011kernel}. Furthermore, a \emph{composition of K-spaced amino-acid pairs} (CKSAAP) is employed to capture a diverse range of peptide sequences, yielding a comprehensive feature vector. 

Motivated by the success of the SRC and the \emph{kernel trick}, we propose in this work to combine polynomial kernel-based \emph{principal component analysis} (PCA) embedding to reduce the feature space dimensions and \emph{synthetic minority oversampling technique} (SMOTE) using $K$-Means \cite{last2017oversampling} to balance the sample space dimension for the construction of the \emph{kernel SRC} model. Details of the proposed approach, including datasets, feature encoding techniques, and classification methods, are furnished in Section \ref{sec:method}. The experimental analysis and discussion of results are provided in Section \ref{sec:result}. The paper is concluded in Section \ref{sec:conclusion}.

\section{Proposed Approach}\label{sec:method}

\begin{figure}[!htb]
\centering
\includegraphics[width=0.66\textwidth]{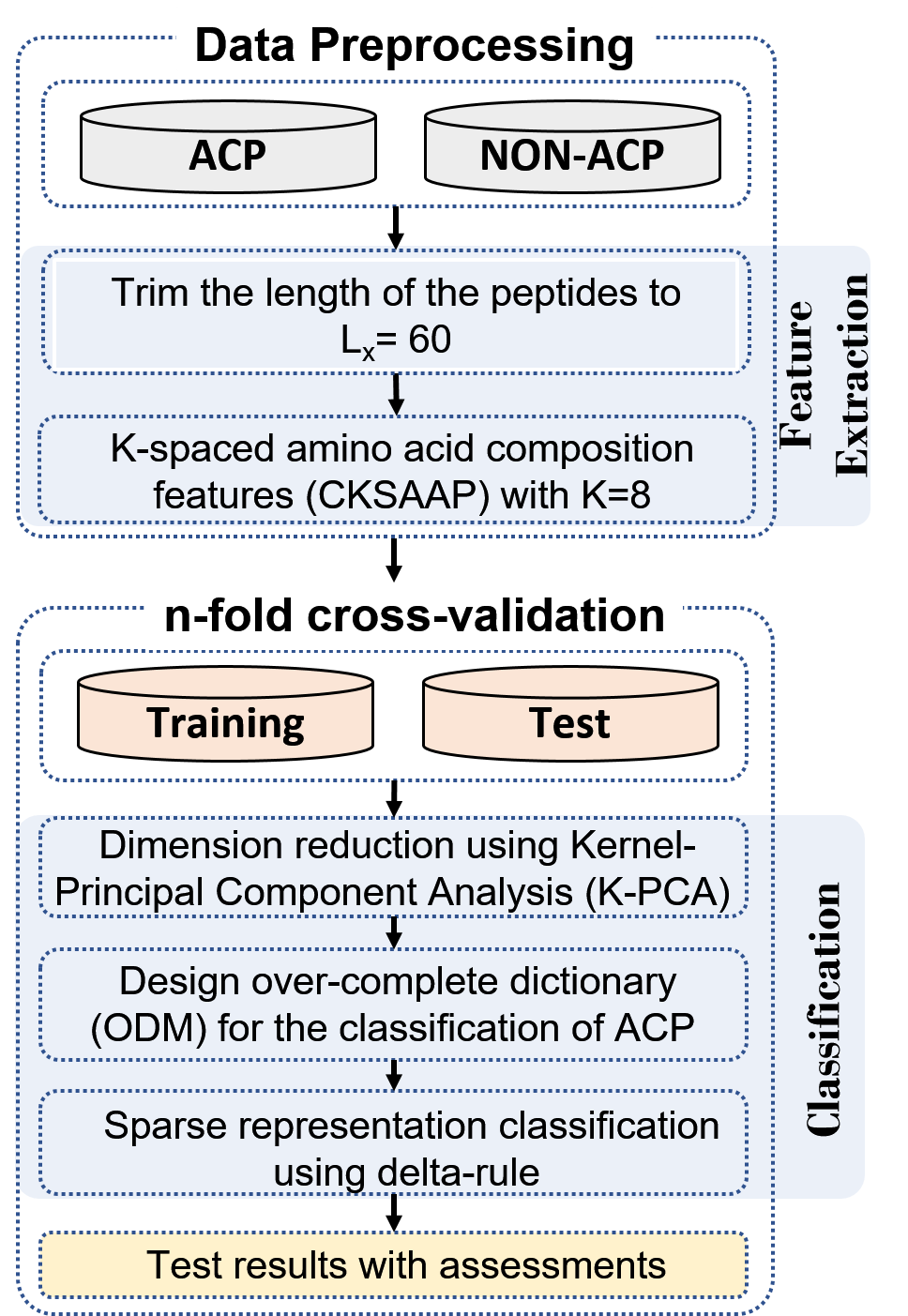}
\caption{Overview of the proposed ACPs classification strategy.}
\label{fig:overview}
\end{figure}

We propose a \emph{kernel sparse representation classification} (KSRC) method in this section which includes feature encoding, dimension reduction for  \emph{OCD matrix} (ODM) design, and $n$-fold cross-validation for model evaluation. Fig. \ref{fig:overview} shows the complete block diagram describing the overall classification process. Individual steps are described in detail in the following subsections. 

\subsection{Dataset} \label{sec:dataset} 

There are many datasets available, including those in \cite{hajisharifi2014predicting, chen2016iacp, wei2018acpred}. Three benchmark datasets are used in this work to design and evaluate the ACP classification strategy. The first dataset, \emph{ACP344}, was obtained from \cite{hajisharifi2014predicting} and contains $344$ peptide sequences, $138$ of which are ACPs and the remaining $206$ are non-ACP samples. The second dataset, \emph{ACP740}, was obtained from previous studies by Chen et al. \cite{chen2016iacp} and Wei et al. \cite{wei2018acpred}). It contains $740$ peptide sequences, $376$ of which are ACP samples and $364$ are non-ACP samples. A filtered and curated version of the ACP740 dataset can be found in \cite{yi2019acp}. Different classifiers are designed and evaluated for each dataset according to the protocols reported in \cite{chen2021acp}. Specifically, $10$-fold cross-validation is used for ACP344, whereas $5$-fold cross-validation is used for ACP740. In the third dataset, two ACP samples were chosen at random from the ACP740 \cite{yi2019acp} dataset, and different mutations were developed for mutation sensitivity analysis. 
It is worth noting that this independent mutant dataset is solely utilized for mutation analysis and is not included in the design of the ODM.

\subsection{Features Encoding}\label{features}

Protein or peptide sequences are often recorded and stored in \emph{FastA} format, with each amino acid represented by an alphabetic symbol; see, for instance, \cite{binz2019proteomics}. These variable-length alphabetic sequences are processed using a variety of sequence encoding techniques, such as AAC, \emph{di-peptide AAC} (DAAC), etc., to extract numerically meaningful features. The AAC is the most basic feature encoding approach, providing a feature vector containing the frequency count of essential amino acids; hence, the overall AAC feature vector length equals the total number of amino acids, i.e., $20$. Similarly, the DAAC is the frequency of peptide pairings, with the total length of the feature vector equal to the number of possible combinations of $20$ amino acid pairs (i.e., $20\times 20 = 400 $). The DAAC feature vector containing the frequencies of $0$-spaced amino acid pairs (i.e., the DAAC of amino acid pairs separated by $K = 0$ residues) is given mathematically by
\begin{align*}
\bpsi_0:=
\begin{bmatrix}
\dfrac{\psi_{A A}}{N_0}& \dfrac{\psi_{A C}}{N_0}& \dfrac{\psi_{A D}}{N_0}& \cdots& \dfrac{\psi_{Y Y}}{N_0}
\end{bmatrix}^T\in\RR^{400}.
\end{align*}
Here, $\psi_{\rm string}$ is the DAAC descriptor furnishing the frequency of the peptide pairing described by the \textit{string} and $N_k:=L_x-(k+1)$ is the number of local sequence windows defined in terms of the protein sequence length $L_x$ and the number of residues $k$ with $0\leq k\leq K$.

\begin{figure*}[!htb]
    \centering 
    \includegraphics[width=1\textwidth]{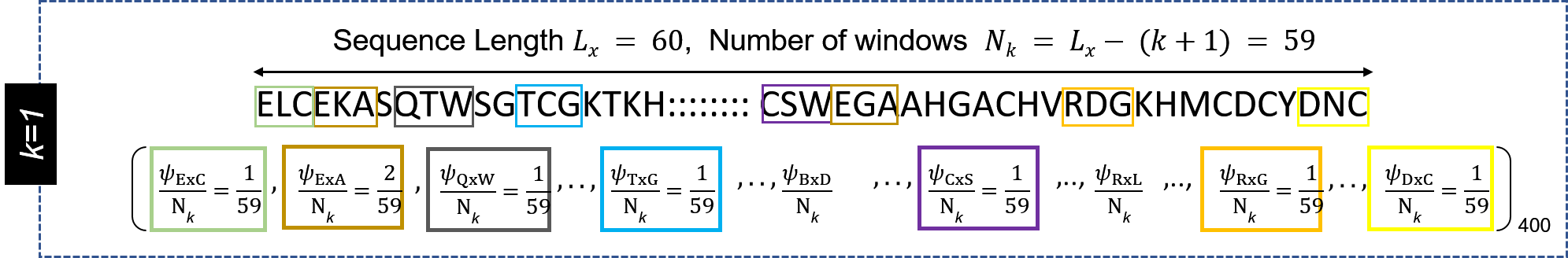}
    \caption{Illustration of k-spaced DAAC $\psi_k$ descriptor calculation for $k=1$. Extracted from \cite{usman2020afp} }
    \label{fig_comp}
\end{figure*}

Both the AAC and DAAC have widely used sequence encoding methods and have been successfully used to design classifiers for various protein and peptide sequences \cite{chen2018ifeature}. However, these techniques are limited in their representation as they do not cover the diverse patterns of the AAC. To improve the pattern capture in DAAC, a modified version is proposed in \cite{chen2018ifeature} by concatenating the DAAC feature vectors of at most $K$-spaced amino acid pairs. For example, for $K=2$, we need to calculate $\bpsi_k$, for $k=0,1,2$, and the final CKSAAP feature vector, $\Psi_{K}$, will be a concatenated version of $\bpsi_0$, $\bpsi_1$, and $\bpsi_2$. Here,
\begin{align*}
\bpsi_1:=&
\begin{bmatrix}
\dfrac{\psi_{A x A}}{N_1}& \dfrac{\psi_{AxC}}{N_1}& \dfrac{\psi_{AxD}}{N_1}& \cdots& \dfrac{\psi_{YxY}}{N_1}
\end{bmatrix}^T\in\RR^{400},
\\
\bpsi_2:=&
\begin{bmatrix}
\dfrac{\psi_{A x x A}}{N_2}& \dfrac{\psi_{A x x C}}{N_2}& \dfrac{\psi_{A x x D}}{N_2}& \cdots& \dfrac{\psi_{Y x x Y}}{N_2}
\end{bmatrix}^T\in\RR^{400},
\\
\Psi_{K}:=&
\begin{bmatrix}
\bpsi_0^T& \bpsi_1^T& \cdots& \bpsi_K^T
\end{bmatrix}^T
\in \RR^{400(K+1)}.
\end{align*}
and $k$ represents a gap value used for the calculation of $k$th DAAC feature vector ($\bpsi_k$) whereas $K$ is the largest possible gap for which CKSAAP feature vector $\Psi_{K}$ is calculated. Fig. \ref{fig_comp} shows an example of the $\bpsi_1$ calculation.

\subsection{Dimensionality Reduction using Kernel PCA}

In machine learning, a large amount of data is often considered useful. A \emph{curse of dimensionality} is nonetheless created when there are few measurements or samples but more attributes (i.e., $A>M$ with $A$ and $M$ being the number of attributes and measurements, respectively). In this study, our dataset has a small number of samples (less than $1,000$) but the description from the CKSAAP with $K=8$ contains $3,600$ attributes. This curse of dimensionality not only makes our classification problem ill-posed mathematically but is also very crucial for the design of an OCD for sparse representation with $A<M$. To that end, we suggest using principal component analysis to prone out the least informative dimensions $f<M=N+P$ from the original feature space of $A=3600$, allowing us to design an ODM of size $f\times M$. Here, $N$ and $P$ are, respectively, the numbers of negative and positive samples in the dictionary.

\begin{figure}[!htb]
\centering
\includegraphics[width=0.66\textwidth]{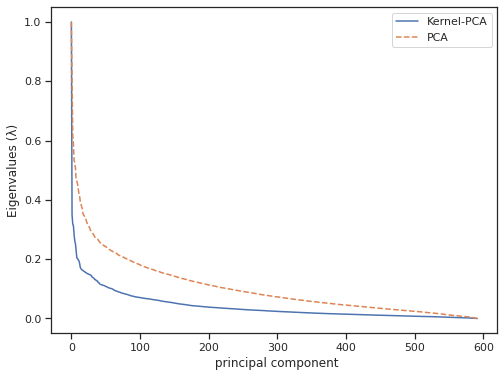}
\caption{Eigenvalues of the top $500$ principal components.}
\label{fig:eigenvalue_PCAvsKPCA_ACP740}
\end{figure}

We employ linear and non-linear projection methods to provide a comparison of the SRC and KSRC methods. A comparison of the eigenvalue spread is presented in Fig.~\ref{fig:eigenvalue_PCAvsKPCA_ACP740} for the ACP740 dataset.  Specifically, from one out of $5$-fold cross-validation results (i.e., $n=5$), the feature set of $A=3600$ attributes, and $M=740\times (n-1)/n = 592$ measurements (samples), is projected in linear and kernel eigenspace and top $600$ eigenvalues are plotted. It can be observed that the KPCA compresses the feature dimension more effectively; hence the relative eigenvalues in the KPCA are smaller. It is also worth noting that the actual number of samples used for the design of the OCD is limited by the training samples in $n$-fold cross-validation. To avoid data leakage, no test sample is used for the estimation of the kernel or for the design of the OCD in our proposed method. First, a PCA projection is learned on training samples, and later the test samples are projected onto the same space using already learned projection.

\subsection{Over-complete Dictionary Matrix for Sparse Representation Classification}
The ODM represents the matrix consisting of feature vectors of ACPs and non-ACPs and is composed of atoms (i.e., training sample vectors). The ODM is used for the SRC in which all ACPs and non-ACPs are characterized using class indices $l=1$ and $l=2$, respectively.

In this section, we take $f$, $N$, and $P$ as the number of features, training samples with negative classes, and training samples with positive classes, respectively.  If $\mathbf{d}_i^{(l)}\in\mathbb{R}^{f}$ represents the $i$th training sample from $l$th class label then the ODM, $\mathbf{D}\in \mathbb{R}^{f\times(N+P)}$, is formed as  
\begin{align*}
\mathbf{D}:=\begin{bmatrix}
\mathbf{d}_1^{(1)} & \mathbf{d}_2^{(1)}& \cdots & \mathbf{d}_N^{(1)} & \mathbf{d}_1^{(2)} & \mathbf{d}_2^{(2)} & \cdots & \mathbf{d}_P^{(2)}\end{bmatrix}.
\end{align*} 
A test sample vector $\mathbf{t}\in \RR^f$ can be represented as
\begin{align*}
\mathbf{t}=\mathbf{D}\bgamma, 
\end{align*}
where the coefficient vector $\bgamma\in\RR^{N+P}$ is defined by 
\begin{align*}
\bgamma:=\begin{bmatrix} \gamma_{1}^{(1)} & \gamma_{2}^{(1)}& \cdots & \gamma_{N}^{(1)}& \gamma_{1}^{(2)}& \gamma_{2}^{(2)}& \cdots &\gamma_{P}^{(2)}\end{bmatrix}^T.
\end{align*}

If the true class label of the test sample $\mathbf{t}$ is $l=1$ then all entries $\gamma_{1}^{(2)}, \gamma_{2}^{(2)}, \cdots,  \gamma_{P}^{(2)}$ should be zero. Similarly, if the true class label is $l=2$ then all entries 
$\gamma_{1}^{(1)}, \gamma_{2}^{(1)}, \cdots,  \gamma_{N}^{(1)}$ should be zero.
According to \emph{sparse reconstruction theory}, if the dictionary $\mathbf{D}$ is given then the sparse vector $\bgamma$ can be recovered \cite{naseem2008sparse,naseem2010sparse}.  In principle, the most sparse $\bgamma$ can be sought as the solution to the optimization problem 
\begin{equation}
\arg \ds{\min_\bgamma} \left\|\bgamma\right\|_{0}\ \ \mbox{subject to}\ \  \mathbf{t}=\mathbf{D}\bgamma,
\label{l0}
\end{equation}
where $\left\|\cdot\right\|_{0}$ is the $l_0$-norm that counts the number of non-zeros entries in the vector. 

The constrained optimization problem \eqref{l0} is non-convex, which makes it difficult to find the optimal vector $\bgamma$.  Several algorithms for recovering the sparse vector $\bgamma$ by solving a \emph{convex relaxation} of the constrained optimization problem \eqref{l0} have been proposed in the literature. To that end, these algorithms make use of the $l_1$-norm to solve the \emph{relaxed} optimization problem 
\begin{equation}
\arg{\min_\bgamma} \left\|\bgamma\right\|_{1}\ \ \mbox{subject to}\ \  \mathbf{t}=\mathbf{D}\bgamma.
\label{equ:l1_prob}
\end{equation}
Some notable techniques for solving the optimization problem \eqref{equ:l1_prob} are the BP \cite{chen1994basis}, OMP \cite{pati1993orthogonal}, and MP \cite{gharavi1998fast}. Among these algorithms, the BP is considered the most robust method as it furnishes the most sparse solution, but its computing cost grows exponentially. The OMP technique can provide a reasonable trade-off between sparsity and computational complexity, but the latter is also very high. On the other hand, MP is faster than the BP and OMP techniques, although its optimality in terms of sparsity is not guaranteed. In the proposed approach, we use the MP as the $l_1$-minimization algorithm because of its suitability for the task at hand.

It should be noted that $\bgamma$ is expected to contain high-value entries corresponding to the columns of $\mathbf{D}$ that are relevant to the class label of the probe $\mathbf{t}$.  This embedded information about the class label of $\mathbf{t}$ can be used to identify $\mathbf{t}$. Let 
\begin{align*}
e_l(\mathbf{t}):=\left\|\mathbf{t}-\mathbf{D}\btheta_l(\bgamma)\right\|_2, \qquad \ l=1,2,
\end{align*}
where the vector $\btheta_l(\bgamma)$ has all zero entries except at the locations corresponding to class $l$ where the value is one.  The decision can be ruled in favor of the class using minimum reconstruction error, i.e., 
\begin{align*}
\mbox{class-label}(\mathbf{t})=\arg {\min_l} \left(e_l (\mathbf{t})\right).
\end{align*}

\subsection{Evaluation Protocol}

The proposed algorithm has been evaluated for \emph{true positive rate} (TPR)  or sensitivity ($S_n$), \emph{true negative rate} (TNR) or 
 specificity ($S_p$), \emph{prediction accuracy} (${\rm Acc}$),  \emph{Matthew's correlation coefficient} (${\rm MCC}$), \emph{balanced accuracy} (${\rm Bal. Acc.}$), \emph{Youden's index} (${\rm YI}$), and \emph{$F1$ Score} defined as 
\begin{align*}
&S_n := \frac{TP}{TP+FN}, 
\\
&S_p := \frac{TN}{TN+FP},
\\
&\text{Acc.} := \frac{TP+TN}{TP+TN+FP+FN},
\\
&{\rm  MCC} :={\frac{TPTN-FPFN}{\sqrt{\Delta}}},
\\
&\text{Bal. Acc.} := \frac{S_n + S_p}{2},
\\
&\text{YI} := S_n + S_p - 1,
\\
&\text{F$1$ Score} := 2* \frac {\text{Precision}*S_n}{\text{Precision} + S_n},
\end{align*}
where $TP$, $FP$, $TN$, and $FN$ indicate the true positive, false positive, true negative, and false negative, respectively. Here, 
\begin{align*} 
&\text{Precision} := \frac{TP}{TP+FP},
\\
&\Delta := (TP+FP)(TN+FN)(TP+FN)(TN+FP).
\end{align*}

\section{Experimental Results} \label{sec:result}

In this section, we perform different experiments to validate our methodology, supporting the selection of various hyper-parameters, solver approaches, embedding strategies, the number of principal components, etc.

\subsection{Comparison of Dictionary Matrices}

The robustness and effectiveness of the ODM are the most critical elements of a sparse representation classifier. We employ principal component embedding of the CKSAAP features to create a useful dictionary.  In particular, the two most frequently used approaches, polynomial-kernel projection and linear projection, are compared. Three comparison criteria are used: 1) the compactness and compressing power of the embedding method, 2) the linear separability of ACPs and non-ACPs in the embedding space, and 3) the classification performance.

\begin{figure}[!htb]
\centering
\includegraphics[width=0.66\textwidth]{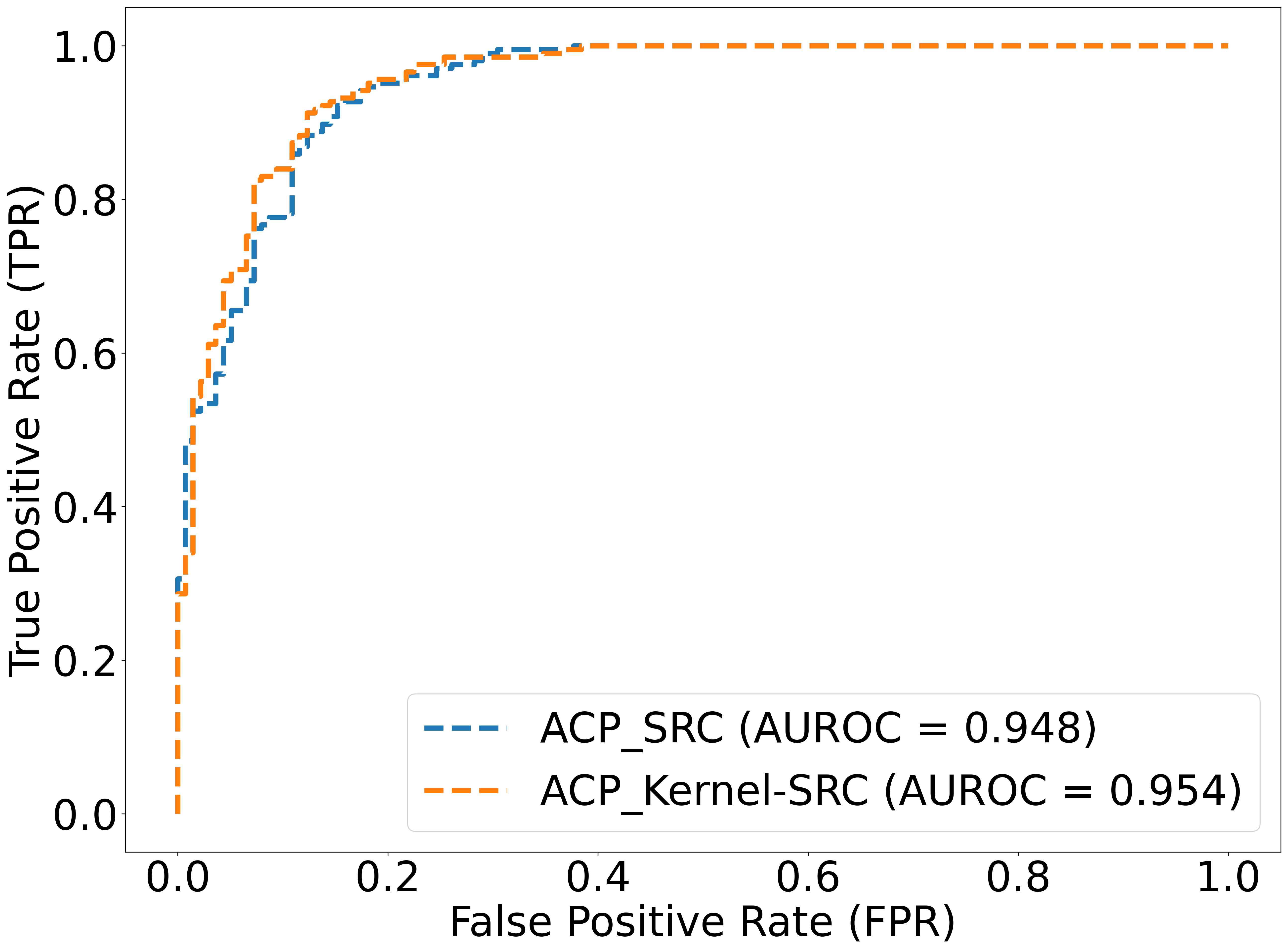}
\caption{Comparison of the AUROC curves of the SRC and KSRC on ACP344 dataset for same configurations using $10$ principal components.}
\label{fig:AUC_acp344_PCA_vs_KPCA}
\end{figure}

As shown in Fig. \ref{fig:eigenvalue_PCAvsKPCA_ACP740}, the kernel PCA requires fewer components to represent the same amount of information as linear PCA. In Fig. \ref{fig:AUC_acp344_PCA_vs_KPCA}, we examine the \emph{area-under-the-receiver-operator-characteristic} (AUROC) curve for the classification of the ACPs from the ACP344 dataset to further substantiate this assertion. The linear PCA-based SRC and polynomial PCA-based KSRC were tested specifically using dictionaries consisting of the first $10$ principal components. The findings show that the KSRC can do better classification with fewer features.

\begin{figure}[!htb]
\centering
\includegraphics[width=0.66\textwidth]{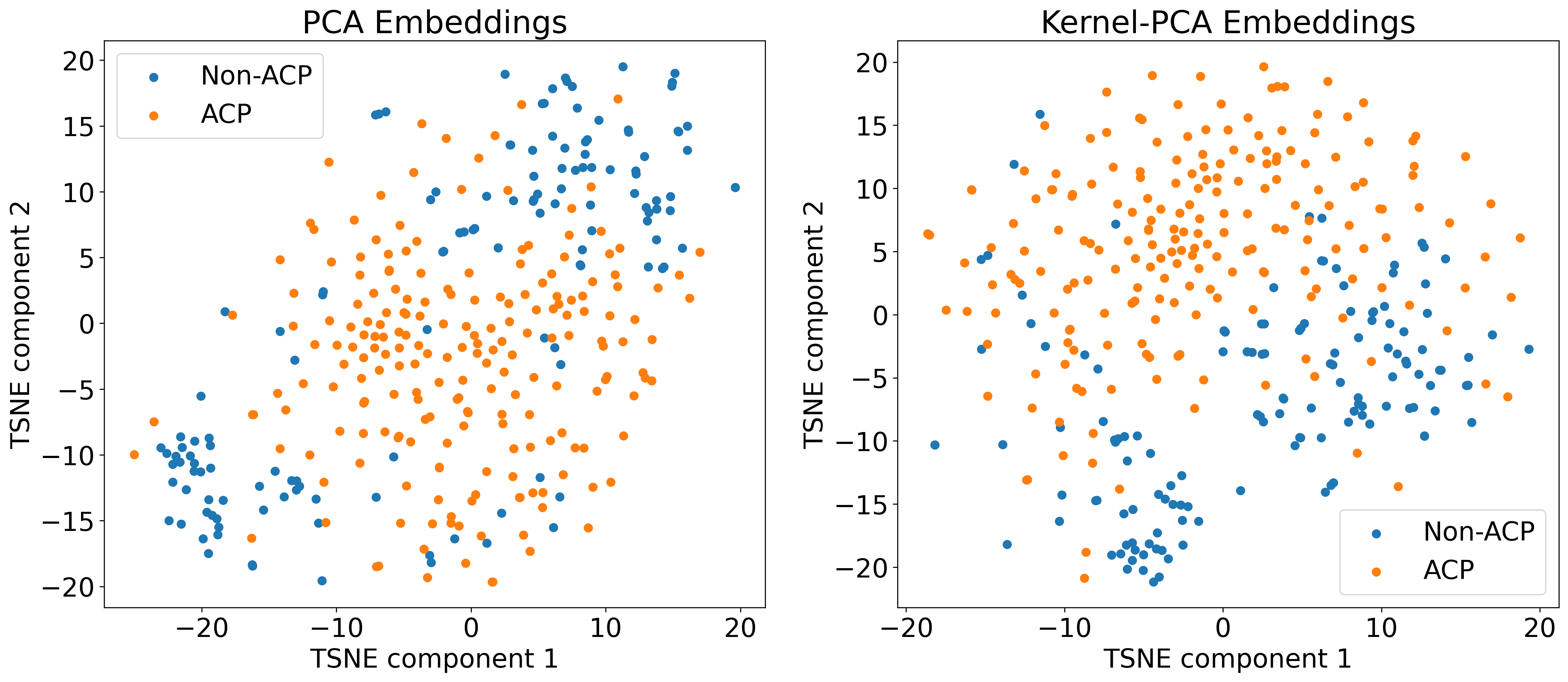}
\caption{Comparison of the linear-PCA and kernel-PCA embedding of the ACP344 dataset using the TSNE and $300$ principal components.}
\label{fig:latent_space_acp344_PCA_vs_KPCA}
\end{figure}

Although compactness is important for sparse representation, the linear separability of class distributions in embedding space is also an important condition. To that end, the $t$-distributed stochastic neighbor embedding (TSNE) \cite{gisbrecht2012linear} plots of linear and kernel PCA embeddings of CKSAAP features of ACP344 dataset are compared in Fig. \ref{fig:latent_space_acp344_PCA_vs_KPCA}. Again, the kernel PCA demonstrates superior linear separability between ACPs and non-ACPs samples.

\begin{figure*}[!htb]
\centering
\includegraphics[width=1\textwidth]{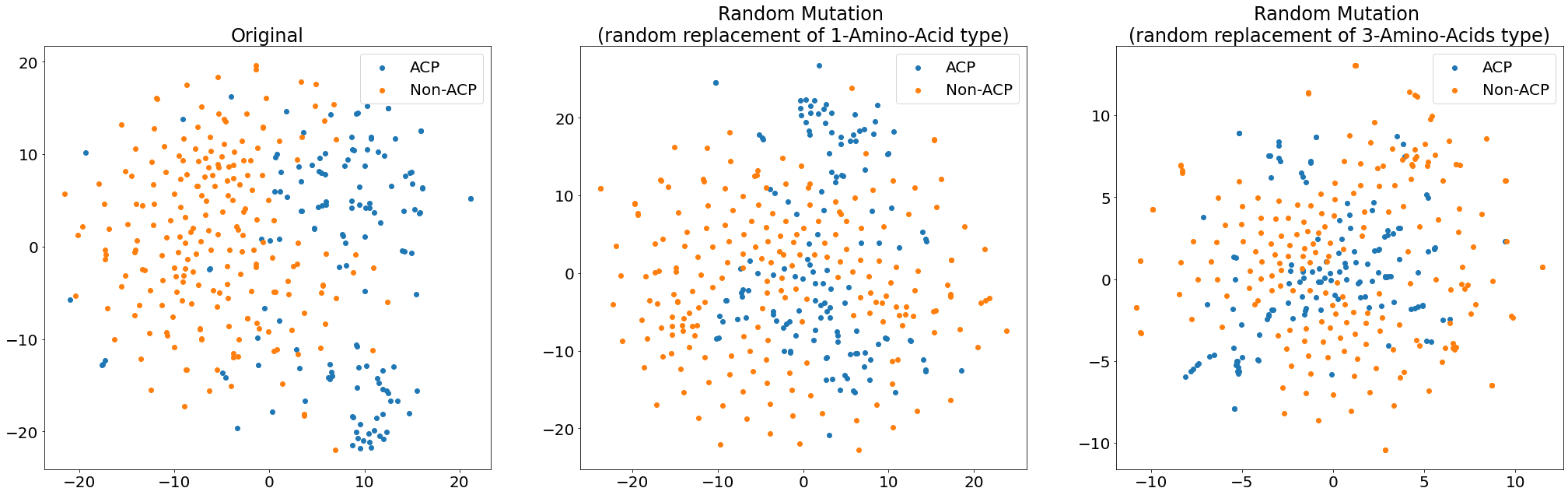}
\caption{Mutation rate and its effect on the feature space. Scatter plot of $2$-components of the TSNE of kernel PCA embedded original and mutant ACPs CKSAAP composition features.}
\label{fig:mutation_acp344_KPCA_TSNE_CKSAAP}
\end{figure*}

In Fig. \ref{fig:mutation_acp344_KPCA_TSNE_CKSAAP}, we compare the variants of the ACP344 dataset to further validate the robustness of the OCD method. In particular, the TSNE plots of the kernel PCA embedding of the CKSAAP features from the original ACP344 dataset and mutants of $138$ ACPs from the ACP344 dataset are compared. The objective of this experiment is to assess the sensitivity of OCD against random mutation. The results show that the separability in the empirical distributions of the ACPs and non-ACPs decreases with the mutation rate. Precisely, when more amino acids in ACPs are mutated, the likelihood of their having anticancer capabilities decreases. A number of statistical methods are available to quantify this distribution separability, ranging from the \emph{strictly standardized mean difference} (SSMD) \cite{zhang2007use} to distribution overlap \cite{park2020gssmd}. However, we are interested in improving classification performance in our experiments, and accordingly, it is the most important aspect of our analysis. 

For the ACP344 dataset on 10-fold cross-validation, the kernel PCA-based KSRC method achieves the maximum mean MCC of $0.8590$ with only $80$ principal components. In contrast, the maximum mean MCC of $0.848$ was achieved in the linear PCA-based SRC with $300$ principal components, which is $1.3\%$ lower than the KSRC performance. In the aforementioned experiments, the MP solver was employed for both the SRC and KSRC methods, while all other settings were unchanged.

\subsection{Comparison of the Optimization Algorithms}

\subsubsection{Classification Performance}

One major challenge in sparse representation classification is to obtain a suitable solution for optimization problem \eqref{equ:l1_prob}. The efficiency of the solution depends on a variety of factors, ranging from the quality of the ODM to the robustness of the solver. There are a variety of algorithms available to deal with ill-posed problems like the one given in \eqref{equ:l1_prob}. The popular $l_1$-minimization algorithms include BP, OMP, and MP. As previously stated, there is a trade-off between the robustness of these algorithms in providing the most sparse solution and their computational efficiency. In the proposed framework, we have adopted the MP algorithm, which is an efficient yet effective $l_1$-minimization algorithm for the task at hand.

To deal with non-linearity and dimension reduction, the polynomial kernel PCA method \cite{scholkopf1997kernel} is used, while the K-means SMOTE \cite{last2017oversampling} is employed to balance the dictionary. The performance of the proposed method is assessed on the benchmark ACP344 dataset for a varying number of principal components. The findings in  Fig.~\ref{fig:mcc_acp344_KPCA} show that the performance of the proposed MP-based KSRC is similar compared to state-of-the-art BP-based KSRC and OMP-based KSRC methods. Specifically, the proposed MP-based ACP-KSRC achieved a mean $10$-fold MCC of $0.8590$ with only $80$ principal components, while the BP and OMP-based approaches achieved a mean $10$-fold MCC of $0.8550$ and $0.8419$ with $40$ and $175$ principal components, respectively. This clearly demonstrates the sparse solution recovery of the BP method. However, due to the nature of our investigation, the sparsity of the solution is not the key aspect. Rather, we are concerned with classification performance, which is superior in the case of MP.
{\begin{figure}[!htb]
\centering
\includegraphics[width=0.66\textwidth]{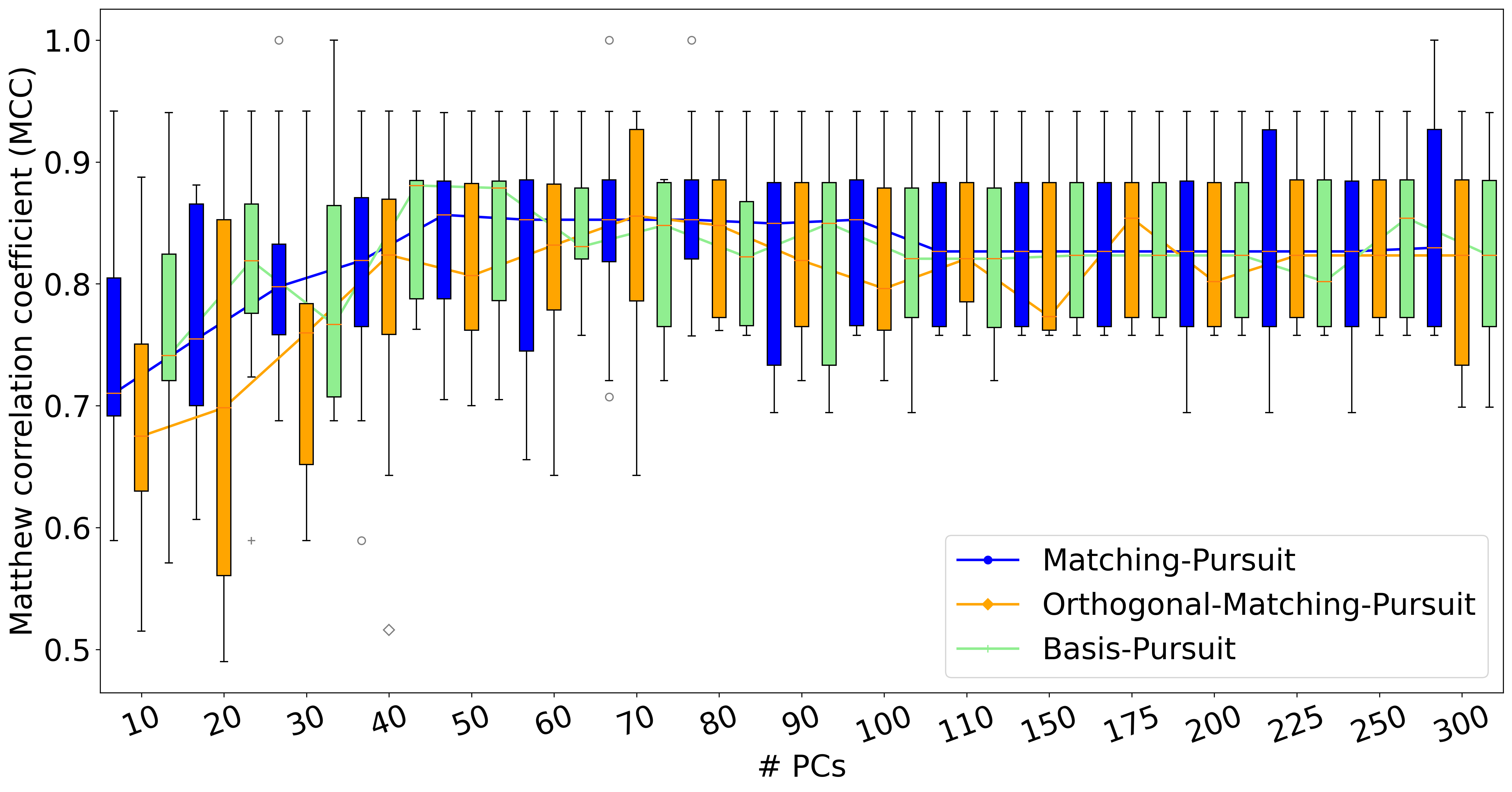}
\caption{Classification performance for different optimization solvers with a varying number of principal components for the ACP344 dataset.}
\label{fig:mcc_acp344_KPCA}
\end{figure}}

\subsubsection{Time Complexity}\label{subsec: Time_Complex}

If we overlook the minor performance gain in MP and examine the run of principal components, we can see that the MP is utilizing more features, which may increase the computational cost. Therefore, we compare the temporal complexity of the preceding experiments for a varying number of principal components. In particular, the \emph{box plots} of all three solvers are shown in Fig.~\ref{fig:recon_time_acp344_KPCA} on a semi-log scale, displaying the median and quartile values for total sample reconstruction time. Interestingly, the time complexity of the MP solver is linearly proportional to the number of principal components, but it is exponential in BP and OMP. This means that even with double the number of principal components, the classification time in the MP is still lower than the BP and equivalent to the OMP with $40$ principal components. The best performance configuration in the OMP is attained with $175$ principal components, the processing time required in the OMP and BP is roughly ten times that of the MP. All experiments were carried out on a freely available \emph{Colab-Notebook} equipped with \emph{Intel Xeon CPU} @2.20 GHz, and $13$ GB RAM.
{\begin{figure}[!htb]
\centering
\includegraphics[width=0.66\textwidth]{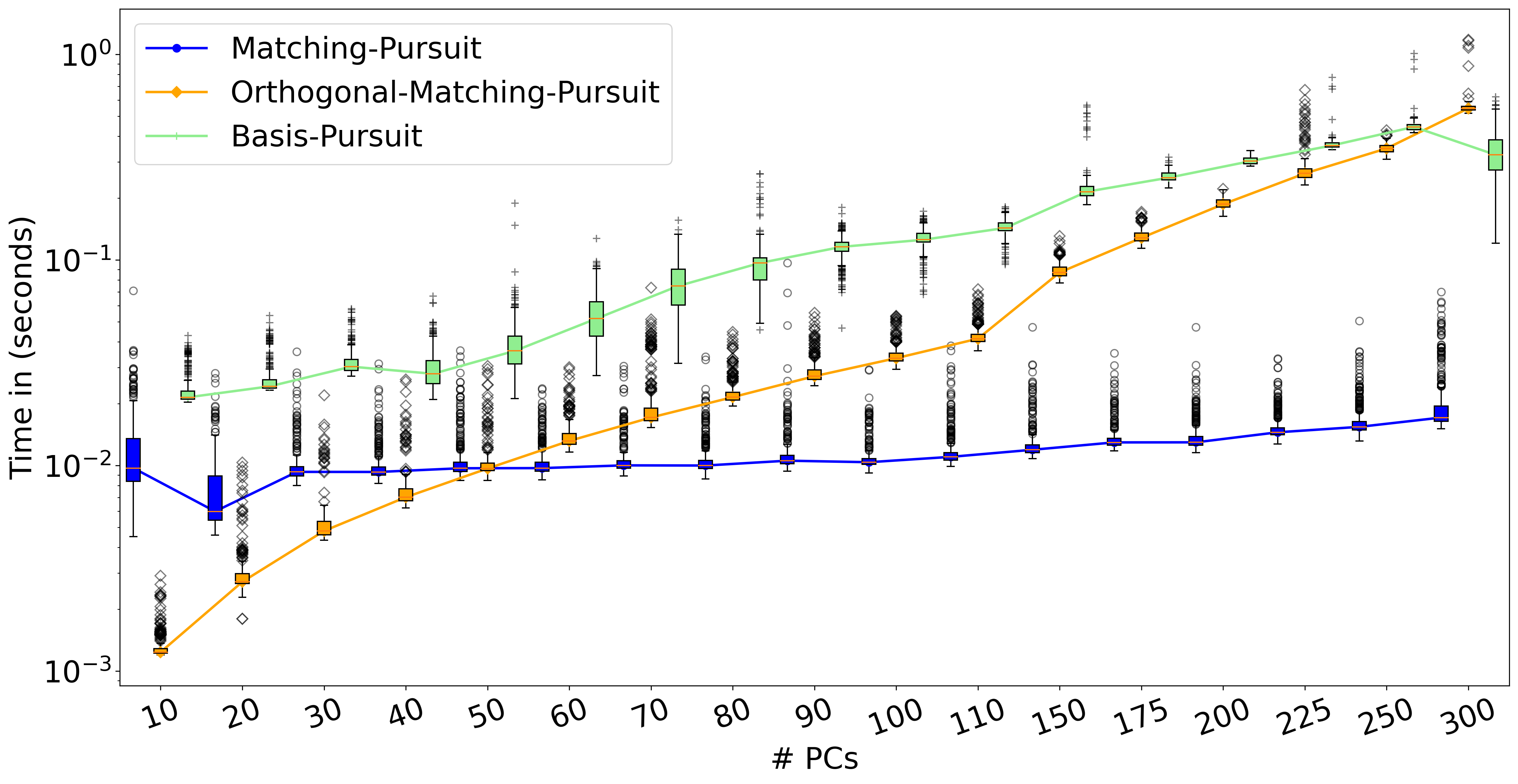}
\caption{Reconstruction time for different configurations and a varying number of principal components for the ACP344 dataset.}
\label{fig:recon_time_acp344_KPCA}
\end{figure}}

\subsection{Comparison with State-of-the-Art ACP Classification Approaches}

In this section, we compare the performance of the proposed ACP-KSRC method with that of the current state-of-the-art ACP classification algorithms on the ACP344 \cite{hajisharifi2014predicting} and ACP740 \cite{yi2019acp} datasets. It should be noted that the proposed ACP-KSRC does not have a training phase, and the training data is only used to construct the dictionary matrix. However, to make a fair comparison, the training and testing samples in all methods were kept consistent as described in previously published research. For instance, the ACP344 dataset is assessed using $10$-fold cross-validation, whereas the ACP740 dataset is evaluated using $5$-fold cross-validation. Other important details about the evaluation of specific datasets are given in the relevant subsections.

\subsubsection{ACP344 Dataset}\label{sec:res_acp344}

In Table \ref{tbl:res_acp344}, we compare the performance statistics of different algorithms on the ACP344 dataset. The number of principal components for the dictionary matrix is set at $f=80$ in this experiment. Since the dataset is unbalanced, the conventional accuracy metric is not a suitable representative of the overall performance. To that end, class-specific evaluation parameters such as the MCC and Youden's index are used to indicate the overall classification ability of the classifier.
Notably, the proposed method yields the best results, demonstrating its ability to effectively differentiate the features of the ACPs. In particular, the ACP-KSRC achieved the highest MCC value of $0.85$ which is $27.06\%$ higher than the ACP-DL, $1.18\%$ higher than the ACP-LDF with the RF and SVM classifiers, $2.35\%$ higher than the ACP-LDF with \emph{LibD3C} classifier, and $4.71\%$ higher than the SAP with the SVM classifier \cite{{ge2020enacp}}. This supports the claim that the proposed method can be used to predict new ACPs or ACP-like peptides. Other assessment metrics mirror this efficacy, indicating the distinct potential between the ACPs and non-ACPs.
\begin{table}[!htb] 
\caption{Performance comparison of the ACP-KSRC with contemporary methods on ACP344 dataset.}
\label{tbl:res_acp344}
	\begin{center} 
		\resizebox{1\columnwidth}{!}
		{\begin{tabular}{lcccccccc}
			\hline
{\bf Methods} & {$S_n$} &  {$S_p$}&  {\bf Acc.}  & {\bf Bal. Acc.} & {\bf MCC} &{\bf YI } & {\bf F1-score}  &{\bf Classifier} \\
\hline
SAP \cite{ge2020enacp}        &  86.23\%  &  95.63\%  & 91.86\% & 90.93\% & 0.83  & 0.81 & 0.89 & SVM  \\ \hline
ACP-LDF \cite{ge2020enacp}        &  87.70\%  &  96.10\% & 92.73\% & 91.90\% & 0.84  & 0.83 & 0.92 & SVM \\ \hline
ACP-LDF \cite{ge2020enacp}        &  85.50\%  &  96.10\% & 92.15\% & 91.05\% & 0.83  & 0.82 & 0.92 & LibD3C  \\ \hline
ACP-LDF \cite{ge2020enacp}        &  86.20\%  &  97.10\% & 92.70\% & 91.65\% & 0.84  & 0.83 & 0.92 & RF  \\ \hline
ACP-DL \cite{yi2019acp}        &  75.82\%  &  86.32\% & 82.16\% & 81.07\% & 0.62  & 0.62 & 0.77 & LSTM  \\ \hline
ACP-KSRC         &  97.07\%  &  86.97\% & 93.02\% & 91.89\% & 0.85  & 0.84 & 0.94 & SRC  \\ \hline
\end{tabular} }
\end{center}
\end{table}

\subsubsection{ACP740 Dataset}\label{sec:res_acp740}

We compare our results for the ACP740 dataset with the ACP-DL \cite{yi2019acp} and ACP-DA\cite{chen2021acp} in Table \ref{tbl:res_acp740}, as these are the only algorithms that used the ACP740 dataset. The proposed method outperforms both algorithms in terms of the class-specific evaluation parameter MCC for $f=100$ principal components. In particular, the ACP-KSRC achieved the highest MCC value of $0.67$ which is $6\%$ and $4.48\%$ higher than the ACP-DL and ACP-DA, respectively. This efficacy is also reflected in other evaluation metrics, indicating the ability of the ACP-KSRC to discriminate between ACPs and non-ACPs. This suggests that the proposed method can be used to predict ACPs or ACP-like peptides.
\begin{table}[!ht] 
\caption{Performance comparison of ACP-KSRC and contemporary methods on ACP-$740$ dataset.}
\label{tbl:res_acp740}
	\begin{center} 
		\resizebox{1\columnwidth}{!}
		{\begin{tabular}{lcccccccc}
			\hline
{\bf Methods} & $S_n$ &  $S_p$ &  {\bf Acc.}  & {\bf Bal. Acc.} & {\bf MCC} &{\bf YI } & {\bf F1-score}  &{\bf Classifier} \\
\hline
ACP-DL \cite{yi2019acp}        &  82.61\%  &  80.59\%  & 83.48\% & 83.3\% & 0.63  & 0.62 & 0.71& LSTM  \\ \hline
ACP-DA \cite{chen2021acp}        &  86.98\%  &  83.26\% & 82.03\% & 85.12\% & 0.64  & 0.70 & 0.85 & MLP  \\ \hline
ACP-KSRC         &  86.23\%  &  81.62\% & 83.91\% & 83.94\% & 0.67  & 0.67 & 0.84 & SRC  \\ \hline

\end{tabular}}
\end{center}
\end{table}


\subsection{Mutation Analysis}
For mutation sensitivity analysis, we have used two ACP samples randomly selected from the ACP740 dataset. Two separate peptides were chosen to have distinct 3D structures, i.e., one had more redundant amino acids than the other. Different mutants of the sequences were constructed for sensitivity analysis. Table \ref{tbl:res_acp740_mutation_10fold} lists the original and mutant sequences.
\begin{table}[!ht] 
\caption{Mutation effect and classification-score sensitivity of ACP-KSRC.}
\label{tbl:res_acp740_mutation_10fold}
	\begin{center} 
		\resizebox{1\columnwidth}{!}
		{\begin{tabular}{l|c|c}
			\hline
{\bf Mutation / (Seq-ID)} & {\bf Sequence} &{\bf Classification-Score} \\
\hline
\hline
Original / (A) &  ALSKALSKALSKALSKALSKALSK & $0.781$  \\ \hline
\hline
Point Mutation / (B) &  ALSKALSKALS\textbf{E}ALSKALSKALSK & $0.762$ \\ \hline
Loop Mutation / (C) &    ALSKALSKALSK\textbf{SQAE}ALSKALSK & $0.759$ \\ \hline
\hline
Original / (D) &  ACDCRGDCFCGGGGIVRRADRAAVP & $0.745$ \\ \hline
\hline
Point Mutation / (E) &  ACDCRG\textbf{K}CFCGGGGIVRRADRAAVP & $0.705$ \\ \hline
Point Mutation / (F) &  ACDCRGDCFCGGGGIVRRA\textbf{K}RAAVP & $0.710$ \\ \hline
Double Point Mutation / (G) &  ACDCRG\textbf{K}CFCGGGGIVRRA\textbf{K}RAAVP & $0.650$ \\ \hline
Loop Mutation / (H) &  ACDCRGDCFC\textbf{SSSS}IVRRADRAAVP & $0.556$  \\ \hline
\end{tabular} }
\end{center}
\end{table}

For a single point mutation analysis, one amino acid from the middle of the peptide sequence is substituted with an amino acid having the opposite property (e.g., non-polar to polar, positive to negative charge, etc.). Similar criteria are employed for double point mutations, but the process is repeated for two amino acids. In loop mutation, multiple peptide composition pairs are replaced with their opposite counterparts. 

For all the sequences given in Table~\ref{tbl:res_acp740_mutation_10fold}, the classification score is predicted using the proposed ACP-KSRC method. The prediction scores were generated using the OCD of ACP740 dataset after $10$-fold cross-validation and each fold sequence from Table~\ref{tbl:res_acp740_mutation_10fold} was used as an independent test sample. Finally, the average prediction score was calculated by taking the mean of the individual fold results. It is noteworthy to point out that the mutant sequence dataset was not used in the design of the OCD matrix.

The prediction score is decreasing for both peptides with higher mutation rates, as predicted. However, the classification score is more sensitive to mutation for sequence (D) as compared to sequence (A). To figure out why this is happening, \emph{Alphafold2} \cite{jumper2021highly} was used to predict the 3D structure of sequences in Table~\ref{tbl:res_acp740_mutation_10fold}. 

\begin{figure}[!htb]
\centering
\includegraphics[width=1\textwidth]{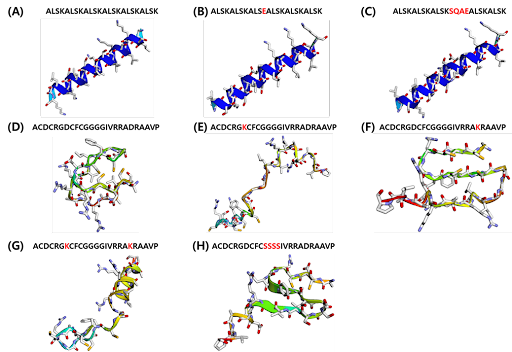}
\caption{\emph{AlphaFold2} predicted 3D structures of original and mutant peptide sequences. (A) and (D) structure of original sequences in Table~\ref{tbl:res_acp740_mutation_10fold}. (B), (E) and (F) structures after point mutation. (G) structure after double-point-mutation. (C) and (H) structures after loop mutation.}
\label{fig:Protein_Structure_AlphaFold}
\end{figure}

It can be observed in Fig.~\ref{fig:Protein_Structure_AlphaFold} that the alpha helix structure of sequence (A) is unaffected by point (B) and loop mutation (C), however, the flexible structure of sequence (D) exhibits a notable difference even with a single point mutation (E). This 3D structure-based mutation analysis demonstrates that the prediction score of the proposed ACP-KSRC is sensitive to structural variation and, hence, can be used as a valuable tool for large-scale ACP screening.

\section{Conclusion}\label{sec:conclusion}

Cancer, as the most challenging disease due to its complexity and heterogeneity, requires multifaceted therapeutic approaches. Anticancer peptides (ACPs) provide a promising perspective for cancer treatment, but their large-scale identification and synthesis require reliable prediction methods. In this study, we have provided an ACP classification strategy which makes use of sparse representation classification combined with kernel principal component analysis (KSRC). The proposed ACP-KSRC approach relies on the well-understood statistical theory of sparse representation classification, unlike the conventional \emph{black box} methods. In particular, we have designed over-complete dictionary matrices using the embedding of the composition of the K-spaced amino-acid pairs (CKSAAP). To deal with non-linearity and dimension reduction, the kernel principal component analysis (KPCA) method is used, while to balance the dictionary, the SMOTE oversampling technique is also utilized. The proposed method is evaluated on two benchmark datasets for well-known statistical parameters and is found to outperform the existing methods. The results indicate the highest sensitivity with the highest balanced accuracy, which can be useful in the understanding of the structural and chemical properties and the development of new ACPs. 

\section*{Competing interests}

All the authors declare that they have no competing interests.

\section*{Acknowledgment}

The authors would like to thank Dr. Shujaat Khan for his valuable suggestions and for providing a Python implementation of his sparse representation classification toolbox. This work was supported by the Nazarbayev University, Kazakhstan through Faculty Development Competitive Research Grant Program (FDCRGP) under Grant 1022021FD2914.

\bibliographystyle{IEEEtran}
\bibliography{References}

\begin{thebibliography}{10}
\providecommand{\url}[1]{#1}
\csname url@samestyle\endcsname
\providecommand{\newblock}{\relax}
\providecommand{\bibinfo}[2]{#2}
\providecommand{\BIBentrySTDinterwordspacing}{\spaceskip=0pt\relax}
\providecommand{\BIBentryALTinterwordstretchfactor}{4}
\providecommand{\BIBentryALTinterwordspacing}{\spaceskip=\fontdimen2\font plus
\BIBentryALTinterwordstretchfactor\fontdimen3\font minus
  \fontdimen4\font\relax}
\providecommand{\BIBforeignlanguage}[2]{{%
\expandafter\ifx\csname l@#1\endcsname\relax
\typeout{** WARNING: IEEEtran.bst: No hyphenation pattern has been}%
\typeout{** loaded for the language `#1'. Using the pattern for}%
\typeout{** the default language instead.}%
\else
\language=\csname l@#1\endcsname
\fi
#2}}
\providecommand{\BIBdecl}{\relax}
\BIBdecl

\bibitem{sung2021global}
H.~Sung, J.~Ferlay, R.~L. Siegel, M.~Laversanne, I.~Soerjomataram, A.~Jemal,
  and F.~Bray, ``Global cancer statistics 2020: Globocan estimates of incidence
  and mortality worldwide for 36 cancers in 185 countries,'' \emph{CA: A Cancer
  Journal for Clinicians}, vol.~71, no.~3, pp. 209--249, 2021.

\bibitem{basith2017expediting}
S.~Basith, M.~Cui, S.~J. Macalino, and S.~Choi, ``Expediting the design,
  discovery and development of anticancer drugs using computational
  approaches,'' \emph{Current Medicinal Chemistry}, vol.~24, no.~42, pp.
  4753--4778, 2017.

\bibitem{kaur2022data}
I.~Kaur, M.~Doja, and T.~Ahmad, ``Data mining and machine learning in cancer
  survival research: An overview and future recommendations,'' \emph{Journal of
  Biomedical Informatics}, p. 104026, 2022.

\bibitem{basak2021comparison}
D.~Basak, S.~Arrighi, Y.~Darwiche, and S.~Deb, ``Comparison of anticancer drug
  toxicities: Paradigm shift in adverse effect profile,'' \emph{Life}, vol.~12,
  no.~1, p.~48, 2021.

\bibitem{yi2019acp}
H.-C. Yi, Z.-H. You, X.~Zhou, L.~Cheng, X.~Li, T.-H. Jiang, and Z.-H. Chen,
  ``Acp-dl: a deep learning long short-term memory model to predict anticancer
  peptides using high-efficiency feature representation,'' \emph{Molecular
  Therapy-Nucleic Acids}, vol.~17, pp. 1--9, 2019.

\bibitem{tyagi2013silico}
A.~Tyagi, P.~Kapoor, R.~Kumar, K.~Chaudhary, A.~Gautam, and G.~Raghava, ``In
  silico models for designing and discovering novel anticancer peptides,''
  \emph{Scientific Reports}, vol.~3, no.~1, pp. 1--8, 2013.

\bibitem{atif2022multi}
S.~M. Atif, S.~Khan, I.~Naseem, R.~Togneri, and M.~Bennamoun, ``Multi-kernel
  fusion for rbf neural networks,'' \emph{Neural Processing Letters}, pp.
  1--25, 2022.

\bibitem{khan2018rafp}
S.~Khan, I.~Naseem, R.~Togneri, and M.~Bennamoun, ``Rafp-pred: Robust
  prediction of antifreeze proteins using localized analysis of n-peptide
  compositions,'' \emph{IEEE/ACM Transactions on Computational Biology and
  Bioinformatics}, vol.~15, no.~1, pp. 244--250, 2018.

\bibitem{usman2020afp}
M.~Usman, S.~Khan, and J.-A. Lee, ``Afp-lse: Antifreeze proteins prediction
  using latent space encoding of composition of k-spaced amino acid pairs,''
  \emph{Scientific Reports}, vol.~10, no.~1, pp. 1--13, 2020.

\bibitem{park2020e3}
S.~Park, S.~Khan, and A.~Wahab, ``E3-targetpred: Prediction of e3-target
  proteins using deep latent space encoding,'' \emph{arXiv preprint
  arXiv:2007.12073 (Retrieved on Dec. 20, 2022)}, 2020.

\bibitem{al2021ecm}
U.~M. Al-Saggaf, M.~Usman, I.~Naseem, M.~Moinuddin, A.~A. Jiman, M.~U.
  Alsaggaf, H.~K. Alshoubaki, and S.~Khan, ``Ecm-lse: Prediction of
  extracellular matrix proteins using deep latent space encoding of k-spaced
  amino acid pairs,'' \emph{Frontiers in Bioengineering and Biotechnology},
  vol.~9, 2021.

\bibitem{usman2021aop}
M.~Usman, S.~Khan, S.~Park, and J.-A. Lee, ``Aop-lse: Antioxidant proteins
  classification using deep latent space encoding of sequence features,''
  \emph{Current Issues in Molecular Biology}, vol.~43, no.~3, pp. 1489--1501,
  2021.

\bibitem{chen2018ifeature}
Z.~Chen, P.~Zhao, F.~Li, A.~Leier, T.~T. Marquez-Lago, Y.~Wang, G.~I. Webb,
  A.~I. Smith, R.~J. Daly, K.-C. Chou \emph{et~al.}, ``ifeature: a python
  package and web server for features extraction and selection from protein and
  peptide sequences,'' \emph{Bioinformatics}, vol.~34, no.~14, pp. 2499--2502,
  2018.

\bibitem{hajisharifi2014predicting}
Z.~Hajisharifi, M.~Piryaiee, M.~M. Beigi, M.~Behbahani, and H.~Mohabatkar,
  ``Predicting anticancer peptides with chou's pseudo amino acid composition
  and investigating their mutagenicity via ames test,'' \emph{Journal of
  Theoretical Biology}, vol. 341, pp. 34--40, 2014.

\bibitem{ge2019identifying}
L.~Ge, J.~Liu, Y.~Zhang, and M.~Dehmer, ``Identifying anticancer peptides by
  using a generalized chaos game representation,'' \emph{Journal of
  Mathematical Biology}, vol.~78, no.~1, pp. 441--463, 2019.

\bibitem{ge2020enacp}
R.~Ge, G.~Feng, X.~Jing, R.~Zhang, P.~Wang, and Q.~Wu, ``Enacp: An ensemble
  learning model for identification of anticancer peptides,'' \emph{Frontiers
  in Genetics}, vol.~11, p. 760, 2020.

\bibitem{chen2021acp}
X.-g. Chen, W.~Zhang, X.~Yang, C.~Li, and H.~Chen, ``Acp-da: Improving the
  prediction of anticancer peptides using data augmentation,'' \emph{Frontiers
  in Genetics}, vol.~12, p. 698477, 2021.

\bibitem{agrawal2021anticp}
P.~Agrawal, D.~Bhagat, M.~Mahalwal, N.~Sharma, and G.~P. Raghava, ``Anticp 2.0:
  an updated model for predicting anticancer peptides,'' \emph{Briefings in
  Bioinformatics}, vol.~22, no.~3, p. bbaa153, 2021.

\bibitem{naseem2017ecmsrc}
I.~Naseem, S.~Khan, R.~Togneri, and M.~Bennamoun, ``Ecmsrc: A sparse learning
  approach for the prediction of extracellular matrix proteins,'' \emph{Current
  Bioinformatics}, vol.~12, no.~4, pp. 361--368, 2017.

\bibitem{li2023multi}
Y.~Li, L.~Hu, and W.~Gao, ``Multi-label feature selection via robust flexible
  sparse regularization,'' \emph{Pattern Recognition}, vol. 134, p. 109074,
  2023.

\bibitem{wright2008robust}
J.~Wright, A.~Y. Yang, A.~Ganesh, S.~S. Sastry, and Y.~Ma, ``Robust face
  recognition via sparse representation,'' \emph{IEEE Transactions on Pattern
  Analysis and Machine Intelligence}, vol.~31, no.~2, pp. 210--227, 2008.

\bibitem{hofmann2008kernel}
T.~Hofmann, B.~Sch{\"o}lkopf, and A.~J. Smola, ``Kernel methods in machine
  learning,'' \emph{The Annals of Statistics}, vol.~36, no.~3, pp. 1171--1220,
  2008.

\bibitem{usman2022afp}
M.~Usman, S.~Khan, S.~Park, and A.~Wahab, ``Afp-src: identification of
  antifreeze proteins using sparse representation classifier,'' \emph{Neural
  Computing and Applications}, vol.~34, no.~3, pp. 2275--2285, 2022.

\bibitem{naseem2010sparse}
I.~Naseem, R.~Togneri, and M.~Bennamoun, ``Sparse representation for speaker
  identification,'' in \emph{2010 20th International Conference on Pattern
  Recognition}.\hskip 1em plus 0.5em minus 0.4em\relax IEEE, 2010, pp.
  4460--4463.

\bibitem{bengio2013representation}
Y.~Bengio, A.~Courville, and P.~Vincent, ``Representation learning: A review
  and new perspectives,'' \emph{IEEE Transactions on Pattern Analysis and
  Machine Intelligence}, vol.~35, no.~8, pp. 1798--1828, 2013.

\bibitem{elad2010sparse}
\BIBentryALTinterwordspacing
M.~Elad, \emph{Sparse and Redundant Representations: From Theory to
  Applications in Signal and Image Processing}.\hskip 1em plus 0.5em minus
  0.4em\relax Springer New York, 2010. [Online]. Available:
  \url{https://books.google.com.pk/books?id=d5b6lJI9BvAC}
\BIBentrySTDinterwordspacing

\bibitem{li2013simultaneous}
S.~Li, Q.~Li, G.~Li, X.~He, and L.~Chang, ``Simultaneous sensing matrix and
  sparsifying dictionary optimization for block-sparse compressive sensing,''
  in \emph{2013 IEEE 10th International Conference on Mobile Ad-Hoc and Sensor
  Systems}.\hskip 1em plus 0.5em minus 0.4em\relax IEEE, 2013, pp. 597--602.

\bibitem{chen1994basis}
S.~Chen and D.~Donoho, ``Basis pursuit,'' in \emph{Proceedings of 1994 28th
  Asilomar Conference on Signals, Systems and Computers}, vol.~1.\hskip 1em
  plus 0.5em minus 0.4em\relax IEEE, 1994, pp. 41--44.

\bibitem{pati1993orthogonal}
Y.~C. Pati, R.~Rezaiifar, and P.~S. Krishnaprasad, ``Orthogonal matching
  pursuit: Recursive function approximation with applications to wavelet
  decomposition,'' in \emph{Proceedings of 27th Asilomar Conference on Signals,
  Systems and Computers}.\hskip 1em plus 0.5em minus 0.4em\relax IEEE, 1993,
  pp. 40--44.

\bibitem{gharavi1998fast}
M.~Gharavi-Alkhansari and T.~S. Huang, ``A fast orthogonal matching pursuit
  algorithm,'' in \emph{Proceedings of the 1998 IEEE International Conference
  on Acoustics, Speech and Signal Processing, ICASSP'98 (Cat. No. 98CH36181)},
  vol.~3.\hskip 1em plus 0.5em minus 0.4em\relax IEEE, 1998, pp. 1389--1392.

\bibitem{zhang2010sparseness}
L.~Zhang and W.~Zhou, ``On the sparseness of 1-norm support vector machines,''
  \emph{Neural Networks}, vol.~23, no.~3, pp. 373--385, 2010.

\bibitem{mandal2016employing}
S.~Mandal and A.~K. Sao, ``Employing structural and statistical information to
  learn dictionary (s) for single image super-resolution in sparse domain,''
  \emph{Signal Processing: Image Communication}, vol.~48, pp. 63--80, 2016.

\bibitem{zhang2011kernel}
L.~Zhang, W.-D. Zhou, P.-C. Chang, J.~Liu, Z.~Yan, T.~Wang, and F.-Z. Li,
  ``Kernel sparse representation-based classifier,'' \emph{IEEE Transactions on
  Signal Processing}, vol.~60, no.~4, pp. 1684--1695, 2011.

\bibitem{khan2017novel}
S.~Khan, I.~Naseem, R.~Togneri, and M.~Bennamoun, ``A novel adaptive kernel for
  the rbf neural networks,'' \emph{Circuits, Systems, and Signal Processing},
  vol.~36, no.~4, pp. 1639--1653, 2017.

\bibitem{last2017oversampling}
F.~Last, G.~Douzas, and F.~Bacao, ``Oversampling for imbalanced learning based
  on k-means and smote,'' \emph{arXiv preprint arXiv:1711.00837 (Retrieved on
  Dec. 20, 2022)}, 2017.

\bibitem{chen2016iacp}
W.~Chen, H.~Ding, P.~Feng, H.~Lin, and K.-C. Chou, ``iacp: a sequence-based
  tool for identifying anticancer peptides,'' \emph{Oncotarget}, vol.~7,
  no.~13, p. 16895, 2016.

\bibitem{wei2018acpred}
L.~Wei, C.~Zhou, H.~Chen, J.~Song, and R.~Su, ``Acpred-fl: a sequence-based
  predictor using effective feature representation to improve the prediction of
  anti-cancer peptides,'' \emph{Bioinformatics}, vol.~34, no.~23, pp.
  4007--4016, 2018.

\bibitem{binz2019proteomics}
P.-A. Binz, J.~Shofstahl, J.~A. Vizca{\'\i}no, H.~Barsnes, R.~J. Chalkley,
  G.~Menschaert, E.~Alpi, K.~Clauser, J.~K. Eng, L.~Lane \emph{et~al.},
  ``Proteomics standards initiative extended fasta format,'' \emph{Journal of
  Proteome Research}, vol.~18, no.~6, pp. 2686--2692, 2019.

\bibitem{naseem2008sparse}
I.~Naseem, R.~Togneri, and M.~Bennamoun, ``Sparse representation for ear
  biometrics,'' in \emph{International Symposium on Visual Computing}.\hskip
  1em plus 0.5em minus 0.4em\relax Springer, 2008, pp. 336--345.

\bibitem{gisbrecht2012linear}
A.~Gisbrecht, B.~Mokbel, and B.~Hammer, ``Linear basis-function t-sne for fast
  nonlinear dimensionality reduction,'' in \emph{The 2012 International Joint
  Conference on Neural Networks (IJCNN)}.\hskip 1em plus 0.5em minus
  0.4em\relax IEEE, 2012, pp. 1--8.

\bibitem{zhang2007use}
X.~D. Zhang, M.~Ferrer, A.~S. Espeseth, S.~D. Marine, E.~M. Stec, M.~A.
  Crackower, D.~J. Holder, J.~F. Heyse, and B.~Strulovici, ``The use of
  strictly standardized mean difference for hit selection in primary rna
  interference high-throughput screening experiments,'' \emph{Journal of
  Biomolecular Screening}, vol.~12, no.~4, pp. 497--509, 2007.

\bibitem{park2020gssmd}
S.~Park, S.~Khan, M.~Moinuddin, and U.~M. Al-Saggaf, ``Gssmd: A new
  standardized effect size measure to improve robustness and interpretability
  in biological applications,'' in \emph{2020 IEEE International Conference on
  Bioinformatics and Biomedicine (BIBM)}.\hskip 1em plus 0.5em minus
  0.4em\relax IEEE, 2020, pp. 1096--1099.

\bibitem{scholkopf1997kernel}
B.~Sch{\"o}lkopf, A.~Smola, and K.-R. M{\"u}ller, ``Kernel principal component
  analysis,'' in \emph{International Conference on Artificial Neural
  Networks}.\hskip 1em plus 0.5em minus 0.4em\relax Springer, 1997, pp.
  583--588.

\bibitem{jumper2021highly}
J.~Jumper, R.~Evans, A.~Pritzel, T.~Green, M.~Figurnov, O.~Ronneberger,
  K.~Tunyasuvunakool, R.~Bates, A.~{\v{Z}}{\'\i}dek, A.~Potapenko
  \emph{et~al.}, ``Highly accurate protein structure prediction with
  alphafold,'' \emph{Nature}, vol. 596, no. 7873, pp. 583--589, 2021.

\end{thebibliography}
\end{document}